\documentclass[prb,reprint,showpacs,onecolumn,superscriptaddress,floatfix]{revtex4-2}

\usepackage[dvipsnames]{xcolor}
\usepackage{tabularx}

\usepackage{bm}
\usepackage{graphicx}
\usepackage{physics}

\def\>{\right\rangle}
\def\<{\left\langle}
\def\be{\begin{equation}}
\def\ee{\end{equation}}
\def\ba{\begin{array}{lll}}
\def\ea{\end{array}}
\def\beq{\begin{eqnarray}}
\def\eeq{\end{eqnarray}}
\begin{document}
\author{Fabio Cavaliere}
    \affiliation{Dipartimento di Fisica, Universit\`a di Genova, Via Dodecaneso 33, 16146 Genova, Italy} 
    \affiliation{CNR-SPIN,  Via  Dodecaneso  33,  16146  Genova, Italy}
\author{Luca Razzoli}
    \affiliation{Center for Nonlinear and Complex Systems, Dipartimento di Scienza e Alta Tecnologia, Universit\`a degli Studi dell'Insubria, via Valleggio 11, 22100 Como, Italy} 
     \affiliation{Istituto Nazionale di Fisica Nucleare, Sezione di Milano, via Celoria 16, 20133 Milano, Italy}
\author{Matteo Carrega}
\affiliation{CNR-SPIN,  Via  Dodecaneso  33,  16146  Genova, Italy}
\author{Giuliano Benenti}
\email{giuliano.benenti@uninsubria.it}
    \affiliation{Center for Nonlinear and Complex Systems, Dipartimento di Scienza e Alta Tecnologia, Universit\`a degli Studi dell'Insubria, via Valleggio 11, 22100 Como, Italy} 
    \affiliation{Istituto Nazionale di Fisica Nucleare, Sezione di Milano, via Celoria 16, 20133 Milano, Italy}
    \affiliation{NEST, Istituto Nanoscienze-CNR, I-56126 Pisa, Italy}
    \author{Maura Sassetti}
    \affiliation{Dipartimento di Fisica, Universit\`a di Genova, Via Dodecaneso 33, 16146 Genova, Italy} 
    \affiliation{CNR-SPIN,  Via  Dodecaneso  33,  16146  Genova, Italy}

\title{Hybrid quantum thermal machines with dynamical couplings}

\date{\today}
\begin{abstract}
SUMMARY: Quantum thermal machines can perform useful tasks, such as delivering power, 
cooling, or heating. In this work, we consider hybrid thermal machines, that can execute 
more than one task simultaneously. We characterize and find optimal working 
conditions for a three-terminal quantum thermal machine, where the working medium is 
a quantum harmonic oscillator, coupled to three heat baths, with two of the couplings 
driven periodically in time. We show that it is possible to operate the 
thermal machine efficiently, in both pure and hybrid modes, and to switch between different 
operational modes simply by changing the driving frequency. Moreover, the 
proposed setup can also be used as a high-performance transistor, 
in terms of output--to--input signal and differential gain.
Due to its versatility and tunability, our model may be of interest
for engineering thermodynamic tasks 
and for thermal management in quantum technologies.
\end{abstract}
\maketitle

\section*{Graphical abstract}
\begin{figure}[ht!]
\includegraphics[width=0.7\linewidth]{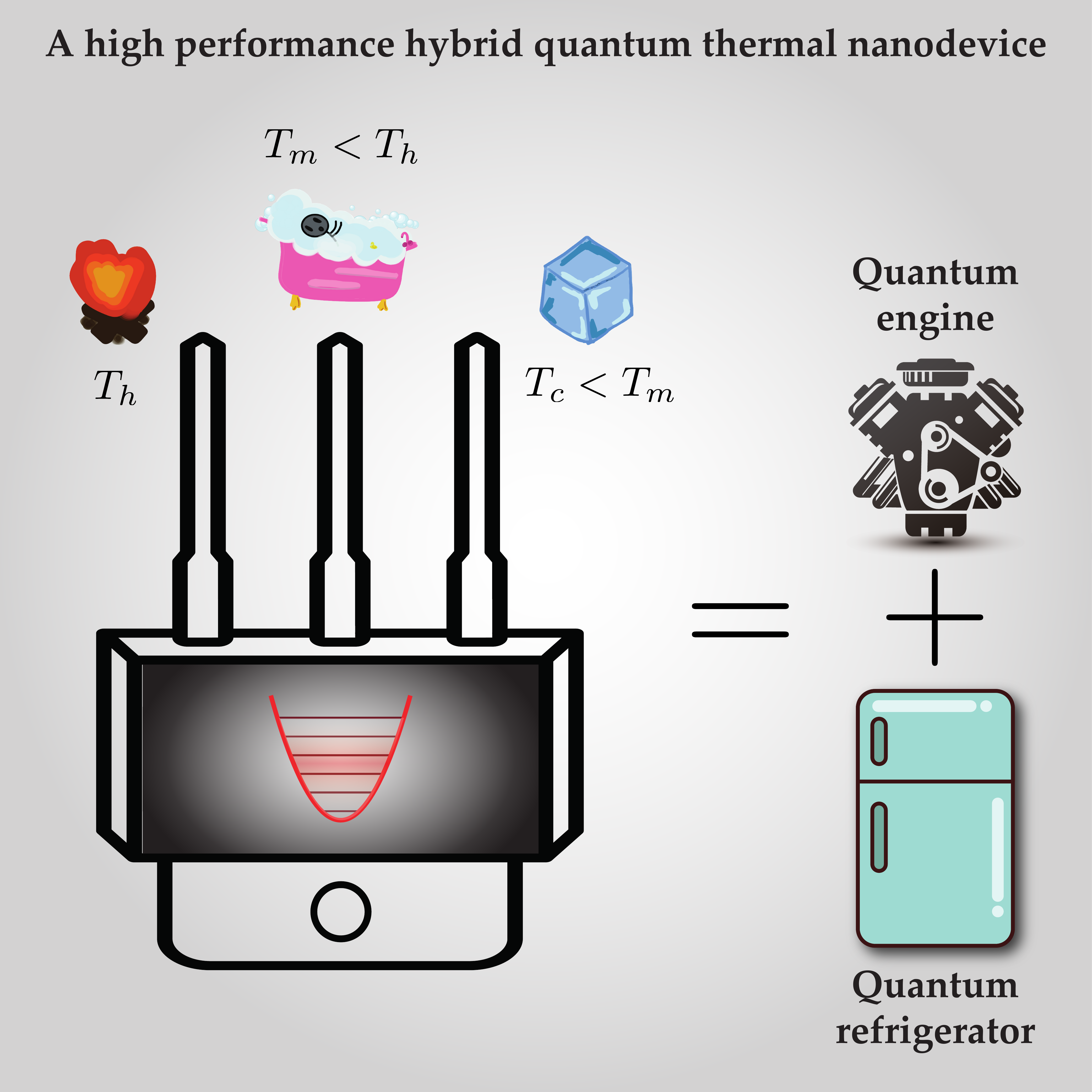}
\end{figure}

\section*{Highlights}

A highly flexible and tunable hybrid thermal machine.

High-performance transistor, in terms of output--to--input signal and differential gain.

\section*{Subject areas}
Physics; Quantum Technologies.
\section{Introduction}
\label{sec:intro}
The fast–paced development of nanoscale technologies is propelling thermodynamics into a new golden era. Just as thermodynamics started in the 1800s spurred by the industrial revolution~\cite{Clausius}, in the same way the miniaturization of devices, and in particular the emergence of new quantum technologies, is pushing the field of thermodynamics into new applied and fundamental challenges~\cite{esposito09,campisi11,kosloff13,kurizki15,sothmann15,anders15,goold16,benenti17,fornieri17,talkner20,paternostro,pekola21,landi22,arracheaR}. Basic questions like the same definitions of work and heat in small systems~\cite{esposito10,talkner20,paternostro}, where quantum mechanics inevitably comes into play, have to be reconsidered with care and are vital to properly characterize the working of nanoscale thermal machines. The minimum temperature achievable in a finite time in small system~\cite{levy12,benenti15,paz1,huber19} is not only a fundamental question related to the third law of thermodynamics but also a practical question for cryogenic applications~\cite{fornieri17,pekola21} and in the initialization of qubits in a quantum computer~\cite{qcbook,qckrantz19, calzona22}. The development of protocols for heat flow control~\cite{giazotto14,pekola21} is the key point to evacuate heat and cool hot spots 
in nanodevices. These are just some of the several fundamental and practical challenges facing thermodynamics.  
More generally, it has been argued~\cite{auffeves22} that a "quantum energy initiative", connecting quantum thermodynamics, quantum information science, quantum physics, and engineering, is needed to develop novel, energy-efficient, sustainable quantum technologies.

There is increasing interest in heat engines where a quantum system is coupled to more than just two reservoirs. For instance, in the cooling by 
heating phenomenon~\cite{pekola07,cleuren12,mari12} the coupling to a third, photonic reservoir, can be used to extract heat from a cold metal, the refrigeration process being powered by absorption of photons. 
It has also been shown that a similar setup, with electron transport 
via a nanostructure bridging two conducting leads and also connected to a third, phonon bath, can be favourable for thermoelectric energy 
conversion~\cite{imry12}. 
Moreover, it has been shown that a third, electronic terminal can improve both the output power and the efficiency at maximum power~\cite{mazza04,sanchez2,blasi21}. In general, 
a multi-terminal device offers enhanced flexibility that might be useful,
for instance, to separate the currents, with charge and heat 
flowing to different reservoirs~\cite{mazza05,sanchez1,three_sanchez}.
Intriguingly, a three-terminal device can work as a hybrid machine, which
performs simultaneously multiple tasks~\cite{imry15,manzano20}, for instance the 
heat from a phonon bath could be used to cool one electron terminal and 
to produce power. On the other hand, three-terminal configurations are
standard in the study of thermal transistors, a device whose development is
essential for an effective thermal management at the 
nanoscale~\cite{casati06,baowen12,joulain16}.

The above studies of multitasking thermal machines considered stationary 
conditions, that is, time-independent Hamiltonians and system-bath couplings. 
On the other hand, driving the system and/or the system-bath couplings
offers enhanced design flexibility, and the possibility to switch 
from some thermodynamic tasks to others simply by tuning the driving frequency.
Here, we consider as a working medium (WM) the paradigmatic model of a quantum harmonic oscillator (QHO), 
which represents a basic building block in several quantum technology 
platforms~\cite{blais04,aspelmeyer, cottet, nori12, arracheaR, calzona22}, 
coupled to three thermal reservoirs. 
We show that, by periodically modulating the couplings in a suitable way
(see below for details and a physical illustration of our model),
it is possible to efficiently run the thermal machine in a pure operation mode
(engine, refrigerator, or heat pump), as well as in a hybrid mode,
combining two of the above. We also show that it is possible to 
switch between different operational modes by changing the 
driving frequency. In addition, we demonstrate that our model can be used as a 
high-performance transistor, in terms of both output--to--input signal 
and of differential gain. Finally, we show that the performance of the 
three-terminal transistor 
is much better than that achievable with the analogous two-terminal device 
with modulated coupling. 

\section{Model and thermodynamic quantities}
\label{sec:model}

\subsection{Three--terminal thermal machine with dynamical couplings}

\begin{figure}[h]
\includegraphics[width=0.8\linewidth]{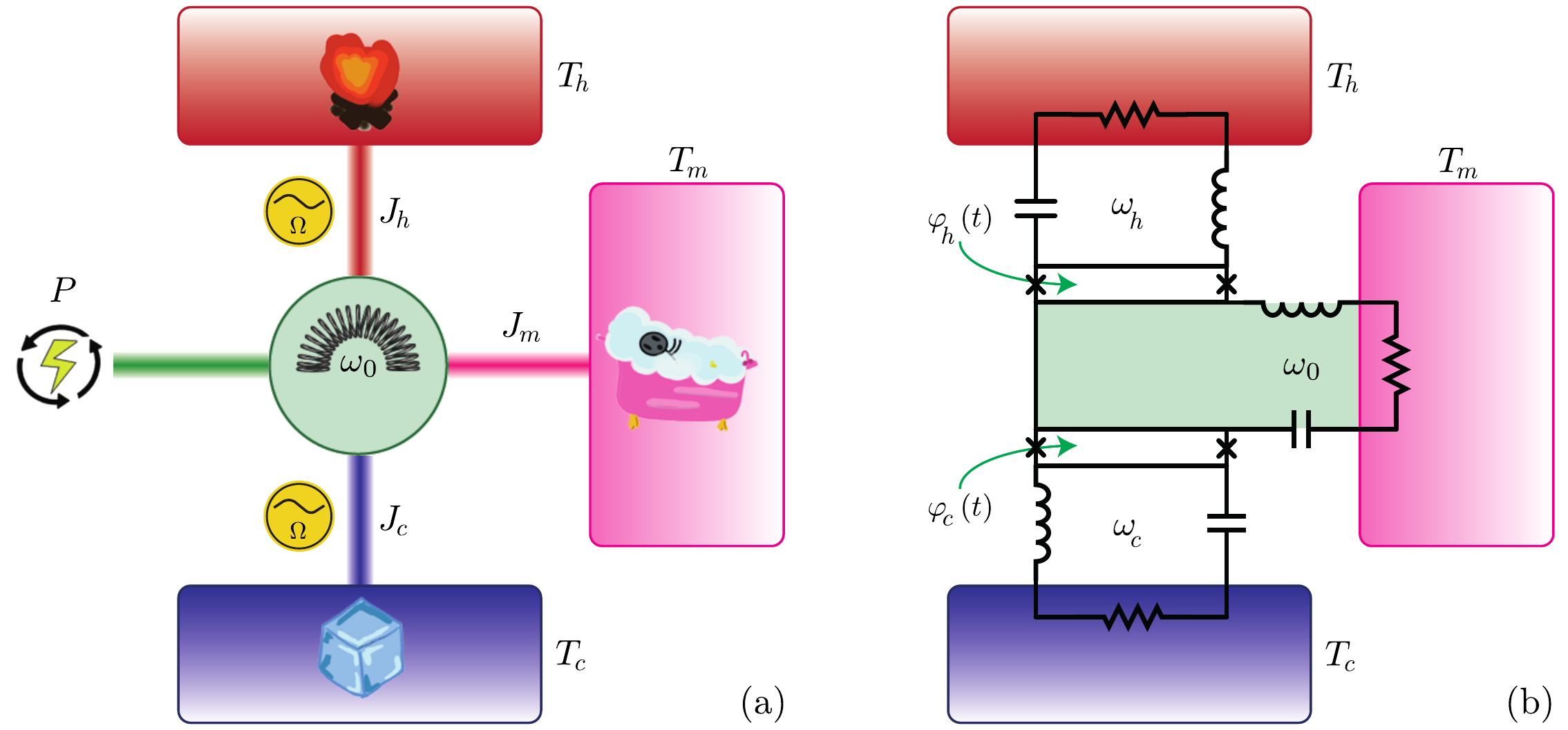}
\caption{{\bf A three--terminal quantum thermal machine}. Panel (a) shows a cartoon of a three-terminal setup consisting of a quantum harmonic oscillator (QHO) in contact with three heat baths at temperatures $T_\nu$ ($\nu=h,m,c$). The coupling with baths $h,c$, is dynamical (driven at frequency $\Omega$), while bath $m$ is statically connected. Here, $J_\nu$ are the energy flows between the QHO and the baths, while $P$ is the total power exchanged via the dynamical couplings. Panel (b) shows a sketch of a possible circuital implementation: the couplings to baths $\nu=h,c$ are implemented in terms of SQUIDs pierced by harmonically driven fluxes $\varphi_{\nu}(t)$ (see text for details).}
\label{fig:setup}
\end{figure}

To study a driven multi-terminal setup that can simultaneously perform several thermodynamic tasks, we focus on the simplest extension beyond the two-terminal paradigm. 
That is, we consider a three--terminal quantum thermal machine, as sketched in Fig.~\ref{fig:setup}(a), where the WM is connected to three reservoirs kept at  temperatures $T_h>T_m>T_c$. Here for 
clarity we have introduced the subscripts $h$ ("hot"), $m$ ("middle") and $c$ ("cold").
The total Hamiltonian is
\be
\label{htot}
H(t) = H_{{\rm WM}} + \sum_{\nu=h,m,c} \big[H_{\nu} + H_{{\rm int},\nu}(t)\big]~.
\ee
where 
\be
H_{{\rm WM}}= \frac{p^2}{2M} + \frac{1}{2}M\omega^2_{0} x^2, 
\ee
with $M$ and $\omega_0$ the mass and the characteristic frequency of the QHO, respectively 
(here and below we set $\hbar=k_{{\rm B}}=1$).
To model the $\nu$-th reservoir we adopt a microscopic description of a thermal bath with many degrees of freedom by using the so-called Caldeira-Leggett framework~\cite{arracheaR, weiss, hu92, CL83, cangemi}: Each reservoir Hamiltonian is described in terms of a collection of independent harmonic oscillators,
\begin{equation}\label{eq:bathsupp}
H_{{\nu}}=  \sum_{k=1}^{\infty} \qty[\frac{P^2_{k,\nu}}{2 m_{k,\nu}} + \frac{m_{k,\nu} \omega^2_{k,\nu }X^2_{k,\nu}}{2}],
\end{equation}
with $X_{k,\nu}$ and $P_{k,\nu}$ the position and momentum operators of the $k$-th oscillator of the $\nu$-th bath with associated mass $m_{k,\nu}$ and frequency $\omega_{k,\nu}$, respectively~\cite{weiss, CL83}. The interaction between the WM and the $\nu$-th bath,
\begin{equation}\label{eq:intsupp}
H_{{\rm int,\nu}}(t)= \sum_{k=1}^{\infty}\left\{-x g_\nu(t)c_{k,\nu}X_{k,\nu} +x^2g_{\nu}^2(t)\frac{c^2_{k,\nu}}{2m_{k,\nu} \omega^2_{k,\nu}}\right\},
\end{equation}
consists of a bilinear coupling between the WM and bath position operators and of the usual counter--term to prevent the renormalization of the WM potential. The interaction strengths are described by the parameter $c_{k,\nu}$, and we have introduced dimensionless functions $g_\nu (t)$ that describe the possible time dependence of the couplings. In particular, we will focus on the situation where the $\nu=h,c$ system/bath couplings are modulated in time~\cite{cangemi, carrega_prxquantum, cavaliere_prr, jurgen}, while the $\nu=m$ coupling is kept constant. Specifically, we assume the two modulations to be in phase, $g_h(t)=g_c(t)=\cos(\Omega t)$, while $g_m(t)=1$. At initial time $t_0=-\infty$, the total density matrix is assumed factorized as $\rho(t_0)=\rho_{{\rm WM}}(t_0)\otimes\rho_h(t_0)\otimes\rho_m(t_0)\otimes\rho_c(t_0)$, 
with $\rho_{{\rm WM}}(t_0)$ the initial WM density matrix, and $\rho_\nu(t_0)$ describing the $\nu$-th bath in thermal equilibrium at temperature $T_\nu$. 

Bath properties can be characterized by the spectral density~\cite{weiss}
\be
\label{eq:spectral}
{\cal J}_\nu(\omega)=\frac{\pi}{2}\sum_{k=1}^\infty \frac{c_{k,\nu}^2}{m_{k,\nu}\omega_{k,\nu}}\delta(\omega-\omega_{k,\nu})\,,
\ee
which encodes memory effects and plays an important role in determining the working regime of 
thermal machines~\cite{cavaliere_prr,breuerrmp,nazir14,groeblacher,illuminatiprl}. Hereafter we choose a strictly Ohmic~\cite{noteOhmic} spectral density ${\cal J}_m(\omega)=M\gamma_m\omega$ for the $\nu=m$ static bath. For the modulated baths $\nu=h,c$, we choose instead a structured environment with a Lorentzian spectral function~\cite{cavaliere_prr, strasberg16, restrepo18, kur21} 
\be
{\cal J}_\nu(\omega)= \frac{d_\nu M\gamma_\nu \omega}{(\omega^2 - {\omega}_\nu^2)^2 + \gamma_\nu^2\omega^2}~\label{eq:lorspd},
\ee
with a peak centered at ${\omega}_\nu$, an amplitude governed by  $d_\nu$, and a damping width determined by $\gamma_\nu$ with $\nu=h,c$. Such structured environments offer a great versatility, i.e. exploiting different resonant conditions between external drive $\Omega$ and $\omega_h$ and $\omega_c$, and can be physically realized in several settings such as quantum cavity~\cite{aspelmeyer, sidebandcooling} or superconducting circuits~\cite{cottet, peropadre, rodrigues, luschmann}. In the latter framework, one can think of a device composed of three superconducting LC circuits, one acting as the WM (with an additional resistive coupling toward the $\nu=m$ reservoir) and the other two as transmission lines coupled to baths via capacitive or inductive couplings~\cite{peropadre, rodrigues}. The temporal modulation can be engineered by varying a mutual capacitance, or by controlling an inductive coupling via external field after embedding the two LC elements into a superconducting quantum interference device (SQUID)-like geometry as in Ref.\cite{peropadre}. An example of such experimental realization is shown in Fig.~\ref{fig:setup}(b).

In this work, we are interested in thermodynamic quantities in the long time limit, when a periodic steady state has been reached. All quantities of interest are thus {\em averaged over one period of the drive} ${\cal T}=2\pi/\Omega$, and are well defined both at weak and strong couplings~\cite{jurgen,esposito19,liu22, brandner166}. The total power and the heat currents, after the quantum ensemble average and averaged over one period of the cycle, read~\cite{carrega_prxquantum, cavaliere_prr}
\beq
&&P=\frac{1}{\cal T}\int_{\bar{t}}^{\bar{t}+{\cal T}}\!\!\!\!\! dt'\,\sum_{\nu=h,c}{\rm Tr}
\Big[\frac{\partial H_{{\rm int},\nu}(t')}{\partial t'}\rho(t_0)\Big]=\frac{1}{\cal T}\int_{\bar{t}}^{\bar{t}+{\cal T}}\!\!\!\!\!dt'\,\sum_{\nu=h,c}{\rm Tr}
\Big[\frac{\partial H_{{\rm int},\nu}}{\partial t'}\rho(t')\Big],
\label{app_PSP}\\
&& J_\nu=-\frac{1}{\cal T}\int_{\bar{t}}^{\bar{t}+{\cal T}}\!\!\!\!\! dt'\,{\rm Tr}
\Big[\frac{d}{dt'}H_{\nu}(t')\rho(t_0)\Big]=-\frac{1}{\cal T}\int_{\bar{t}}^{\bar{t}+{\cal T}}\!\!\!\!\! dt'\,{\rm Tr}
\Big[H_{\nu}\frac{d}{dt'}\rho(t')\Big],\label{app_JSP}
\eeq
where $P$ represents the total power associated to the time evolution of the system/bath couplings $\nu=h,c$ -- with $P>0$ when work is performed on the WM -- and $J_{\nu}$ is the heat current from the  $\nu$-th reservoir, with $J_{\nu}>0$ when the flow is directed towards the WM, see Fig.~\ref{fig:setup}(a). We remark here that  only the two driven couplings $\nu=h,c$ contribute to the power, since $g_m(t) \equiv 1$. 

It is important to notice that the average total power is fully balanced by the average heat currents
\be
 \label{firstlaw}
P +\sum_{\nu=h,m,c} J_\nu=0\,,
\ee
reflecting the first law of Thermodynamics~\cite{benenti17, paternostro}.
Another important quantity to assess thermodynamic performance is the  time average entropy production rate, which can be linked to the heat currents by~\cite{benenti17, esposito19, paternostro}
\be
 \label{entropy}
\dot S=-\sum_{\nu=h,m,c}\frac{J_\nu}{T_\nu}\,,
\ee
and in terms of which the second law of thermodynamics can be formulated as $\dot{S}\ge 0$. Note that the averaged contribution of the WM to the entropy production vanishes as the system's steady state is periodic with the period of the drive.

\subsection{Operating modes of a three-terminal thermal machine}
\label{sec:operatingmodes}

\begin{figure}[h]
\includegraphics[width=0.8\linewidth]{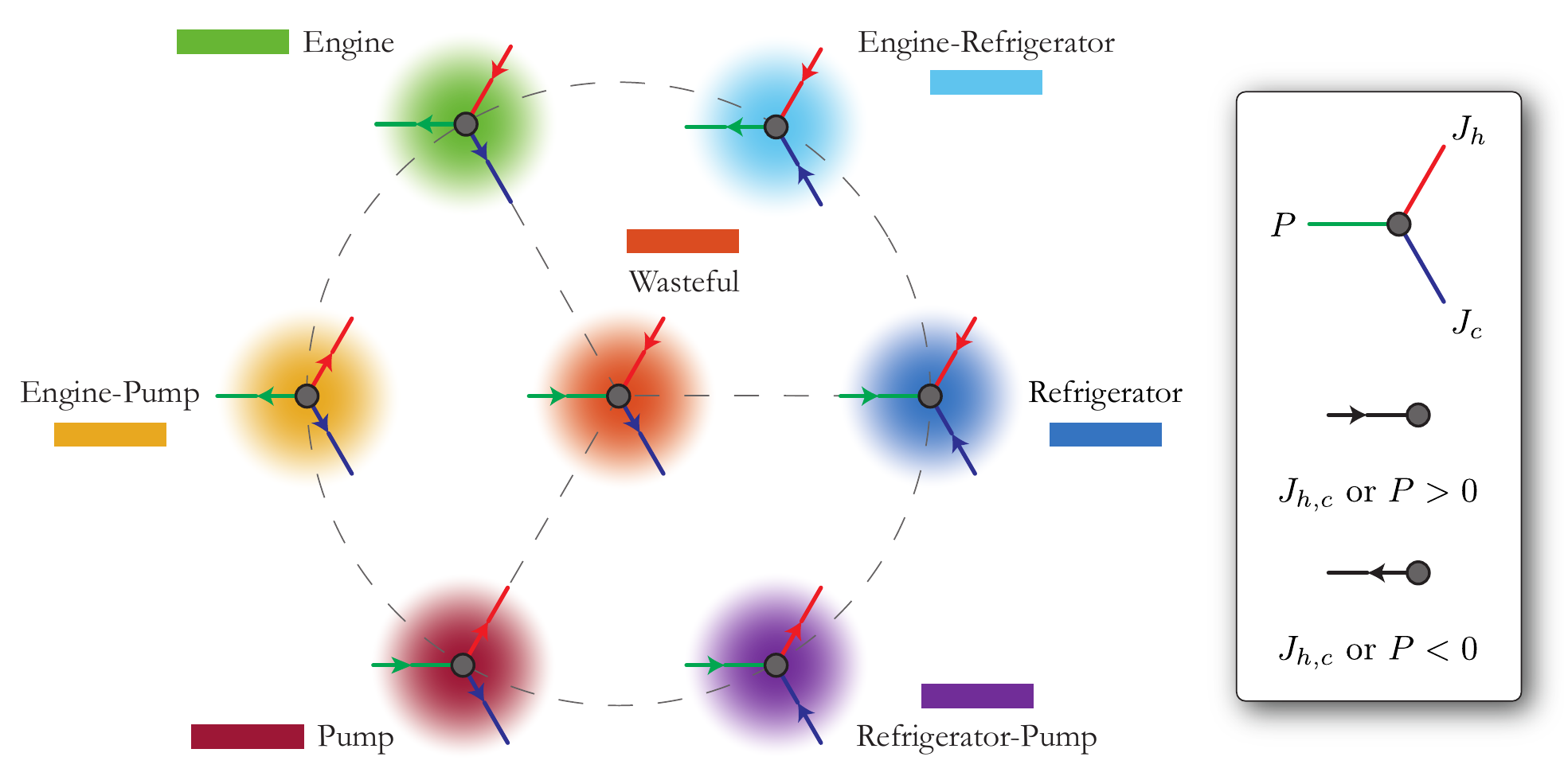}
  \caption{{\bf Operating modes of a driven three--terminal device}. Schematic depiction of all the possible operating modes of a three--terminal device,
  as told by the sign of $J_h$, $J_c$ and $P$. The dashed lines represent the modes that are connected via the reversal of only one among $J_{h,c}$ or $P$. The arrows represent the direction of the latter: the relationship with respect to the sign of currents and power is shown in the legend.}\label{fig:modes}
  \label{table} 
\end{figure}

The operating modes of a thermal machine can be classified studying the sign of thermal currents and total power. Due to the energy conservation constraint,  $P + J_h+J_c+J_m=0$,
we can express the 
entropy production rate $\dot{S}$ in terms of three quantities alone, say $J_h$, $J_c$ and $P$,
as
\be
\dot{S}=-\sum_{\nu=h,m,c}\frac{J_\nu}{T_\nu}= \frac{P}{T_m} + \frac{J_c}{T_m}\left(1-\frac{T_m}{T_c}\right) +\frac{J_h}{T_m}\left(1- \frac{T_m}{T_h}\right)\,.\label{eq:Sdot2}
\ee
There are $2^3$ conceivable operating modes: one however is not physical since the combination $J_h<0$, $J_c>0$ and $P<0$ violates $\dot{S}\geq 0$ -- see Eq.~(\ref{eq:Sdot2}) -- and never occurs. The other modes are schematically depicted in Fig.~\ref{fig:modes}. One is commonly labeled as "wasteful"~\cite{hajiloo, manzano20} since in this configuration $J_c<0$, $J_h>0$ and $P>0$ -- meaning that heat flows from the bath at $T_h$ to the one at $T_c$ while the WM absorbs power. However, this configuration can be a resource for the operation as a thermal transistor, as will be shown in Sec.~\ref{sec:tranny}. The other six operating modes consist of three "pure" ones (engine, refrigerator, heat pump) and three "hybrid" modes~\cite{manzano20, lu22, lopez22}. The existence of the latter implies that a three--terminal device can simultaneously perform multiple thermodynamic tasks, such as employing heat from the hot reservoir to produce power (engine) and simultaneously "lift" heat from the cold reservoir (refrigerator), a possibility out of reach for conventional two--terminal thermal machines~\cite{imry15,arracheaR, manzano20, lu22, lopez22}.

To quantify the performance of a multi-terminal thermal machine in a unified fashion, including also multitasking configurations, it is useful to introduce the so-called input--output exergy efficiency, also known as the second-law efficiency or rational efficiency~\cite{manzano20, lu22, lopez22}. To this end, we split the entropy production rate $\dot{S}= \dot{S}^{(+)}+\dot{S}^{(-)}$ into positive (+) and negative (-) contributions and introduce the exergy efficiency as
\be
\label{phi}
\phi=-\frac{\dot{S}^{(-)}}{\dot{S}^{(+)}}\,,
\ee
where negative (positive) contributions of the entropy production rate appear in the numerator (denominator). Clearly, since $\dot{S} = \dot{S}^{(+)} - | \dot{S}^{(-)} | \geq 0$, one
has $| \dot{S}^{(-)} | \leq \dot{S}^{(+)}$, which implies that $0 \leq
\phi \leq 1$. {Note that, in simpler cases (e.g. a two--terminal setup
operating as an engine), $\phi$ reduces to the standard thermodynamical
efficiency $\eta$ normalized to the relevant Carnot limit~\cite{manzano20}.}
In our case, using Eq.~(\ref{eq:Sdot2}), we can write explicitly
\be
\phi=-\frac{P\vartheta(-P) + J_c (1-T_m/T_c)\vartheta(J_c) + J_h(1-T_m/T_h)\vartheta(-J_h)}{P\vartheta(P) + J_c(1-T_m/T_c)\vartheta(-J_c) +J_h(1-T_m/T_h)\vartheta(J_h)},
\ee
where $\vartheta(x)$ is the step function.

\section{Results}
\label{sec:results}

We will now present typical results which illustrate the versatility of a three--terminal device, demonstrating in particular its ability to perform hybrid tasks.
In addition, we will show that the same setup can be configured as a useful thermal transistor~\cite{dutta, pekola21, gubaydullin, campeny19, gupt, ligato}. Hereafter, we consider the quantum regime~\cite{pekola21}
\be
\omega_0> T_h> T_m> T_c\,.
\ee
We focus on the perturbative regime in which the coupling with the driven Lorentzian baths is much smaller than the one with the static Ohmic bath. 
The amplitude $d_{\nu}$ (for $\nu=h,c$) in Eq.~(\ref{eq:lorspd})
can be expressed in terms of the dimensionless parameter
\begin{equation}
\kappa_{\nu}=\frac{d_{\nu}}{\omega_{\nu}^2\omega_0^2}\ll 1.
\end{equation}
As for the other parameters of the Lorentzian baths, we  assume that their resonant frequencies $\omega_{h,c}>0$ can be independently tuned, with the detuning parameter $\Delta= \omega_h-\omega_c$, and that the widths $\gamma_h=\gamma_c$. Finally, we address the underdamped regime with $\gamma_{\nu}\ll\omega_0$, in which the machine can display its best performances. In these conditions compact analytical expressions can be obtained~\cite{carrega_prxquantum, cavaliere_prr} for the thermal currents $J_{\nu}$ and power $P$: they are reported in Appendix~\ref{app:1}, see Eqs.~(\ref{eq:curpert},\ref{eq:curpow}).

\subsection{Operation as a hybrid device}

\label{sec:mmode}

\begin{figure}[h]
\includegraphics[width=0.8\linewidth]{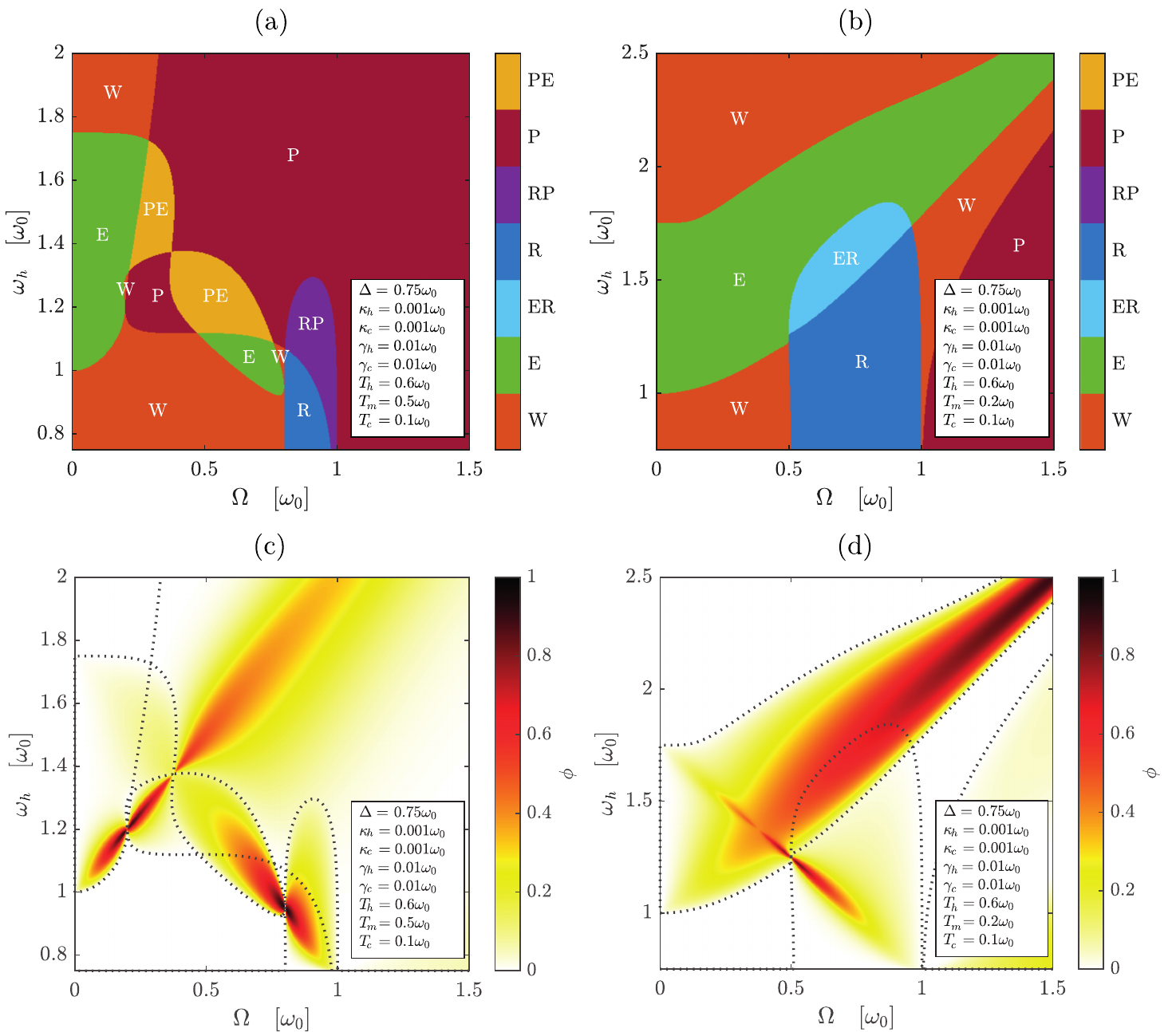}
\caption{{\bf Hybrid operation of a three--terminal device}. Operating modes (a,b) and corresponding exergy efficiency $\phi$ (c,d) as a function of the driving frequency $\Omega$ and $\omega_h$ for a three--terminal quantum thermal machine operating with same parameters except for $T_m=0.5\omega_0$ in panels (a,c) and $T_m=0.2\omega_0$ in panels (b,d). Here we recall that $\Delta=\omega_h-\omega_c$.}
  \label{fig:multimode} 
\end{figure}

The three--terminal quantum device proposed here can display several different operating modes. To maximize their number, we have found that choosing $\Delta>0$ is a key ingredient
(see below for an explanation of this observation).
Figures~\ref{fig:multimode} (a) and (b) show the operating mode of the device as a function of $\Omega$ and $\omega_h$ with a large detuning $\Delta=0.75\omega_0$, for two representative values of the middle temperature $T_m$. When $T_m\sim T_h$ (panel a) we encounter a very rich scenario characterized by all the pure modes along with the refrigerator--pump and pump--engine hybrid behaviours. It is important to stress that this last regime
cannot be observed with only two terminals, see Appendix~\ref{app:2}. Considering instead $T_m\sim T_c$, exemplified by panel (b), we still observe a varied (but somewhat simpler) scenario, which however features the engine--refrigerator hybrid operation, which again is lacking in the two-terminal device. A very intriguing feature of our system is its ability to seamlessly switch between several different modes of choice by tuning $\omega_h$, $\Delta$ and, more simply, using the external driving frequency $\Omega$ as a control knob. As an example, considering the situation in panel (a) and $\omega_h\approx1.5\omega_0$ it is in principle possible to switch back and forth between engine, a hybrid engine-heat pump and a heat pump by sweeping $\Omega$ in a range $\lesssim 0.5\omega_0$. Similar configurations are of course possible by choosing different regions of the parameter space. Importantly, most of the configurations are quite wide with respect to $\omega_h$ and do not depend sensibly on the specific value of $\Delta$, which suggests that the operation and the tunability of device are quite stable and robust.\\
Figures~\ref{fig:multimode} (c) and (d) show the exergy efficiency $\phi$ as a function of $\Omega$ and $\omega_h$ for the same parameters of panels (a) and (b) discussed above. Clearly, $\phi=0$ in the wasteful regions. Away from them, the exergy efficiency can attain significant values $\gtrsim 0.5$ both in the pure as well as in the hybrid modes. One feature clearly stands out: the largest values of $\phi$ occur around two precise lines in the $\Omega$, $\omega_h$ plane. This is because the three--terminal device can be seen as two two--terminal devices in parallel: the machine $\mathcal{M}_1$ operating between $T_m$ and $T_h>T_m$, and $\mathcal{M}_2$ working between $T_c$ and $T_m>T_c$. The behaviour of such a two--terminal device has been analyzed previously~\cite{cavaliere_prr}, 
see Appendix~\ref{app:2} for a brief summary. In particular, the two high-efficiency lines mentioned above correspond to the resonances $\omega_h=\omega_0+\Omega$, due to the operation of $\mathcal{M}_1$ and $\omega_c=\omega_0-\Omega=\omega_h-\Delta$, due to the operation of $\mathcal{M}_2$. This observation also allows us to understand the richness of panels (a) and (b), due to the superposition of the different working modes of the "elementary" two--terminal devices in terms of which the three--terminal one can be interpreted. Such superposition occurs if $\omega_c<\omega_h$ which implies $\Delta>0$, which reveals why this is a favourable regime to observe hybrid operating modes.

\subsection{Operation as a thermal transistor}
\label{sec:tranny}

Three--terminal quantum thermal devices have been envisioned as possible templates to implement a quantum thermal transistor~\cite{hajiloo, pekola21, joulain16, gupt}, i.e. a device in which a small modulation of an 
energy flow (a "input", analogous to the gate/base of an electronic transistor) produces a large modulation of another, larger energy flow (a "output", similar for instance to the source/emitter in a conventional transistor)~\cite{gupt}. Typical setups considered in literature consist of several two--level--systems (TLS) statically coupled to external baths~\cite{joulain16}. There, the modulation of the "input" flow is provided by a modulation of the temperature of the "input" bath (typically, corresponding to the one at temperature $T_m$). 
The periodic modulation of one of the TLSs has been also considered~\cite{gupt}. Proposals of a thermal transistor in a two--terminal device with parametric excitation 
have also been put forward~\cite{campeny19}: the two terminals serve the purpose of source/emitter and drain/collector, while the exchanged power with the parametric actuator serves as the gate/base. In such a transistor the modulation of the gate is achieved via the modulation of the frequency of the parametric excitation, which offers the practical advantage of being simpler to achieve in a precise manner in comparison to the modulation of the temperature of a bath.
In this work we endorse this idea, and  show that a three--terminal quantum thermal transistor with remarkable performances and stability can be achieved in our setup, which outperforms an analogous device with only two terminals.

\begin{figure}[h]
\includegraphics[width=0.8\linewidth]{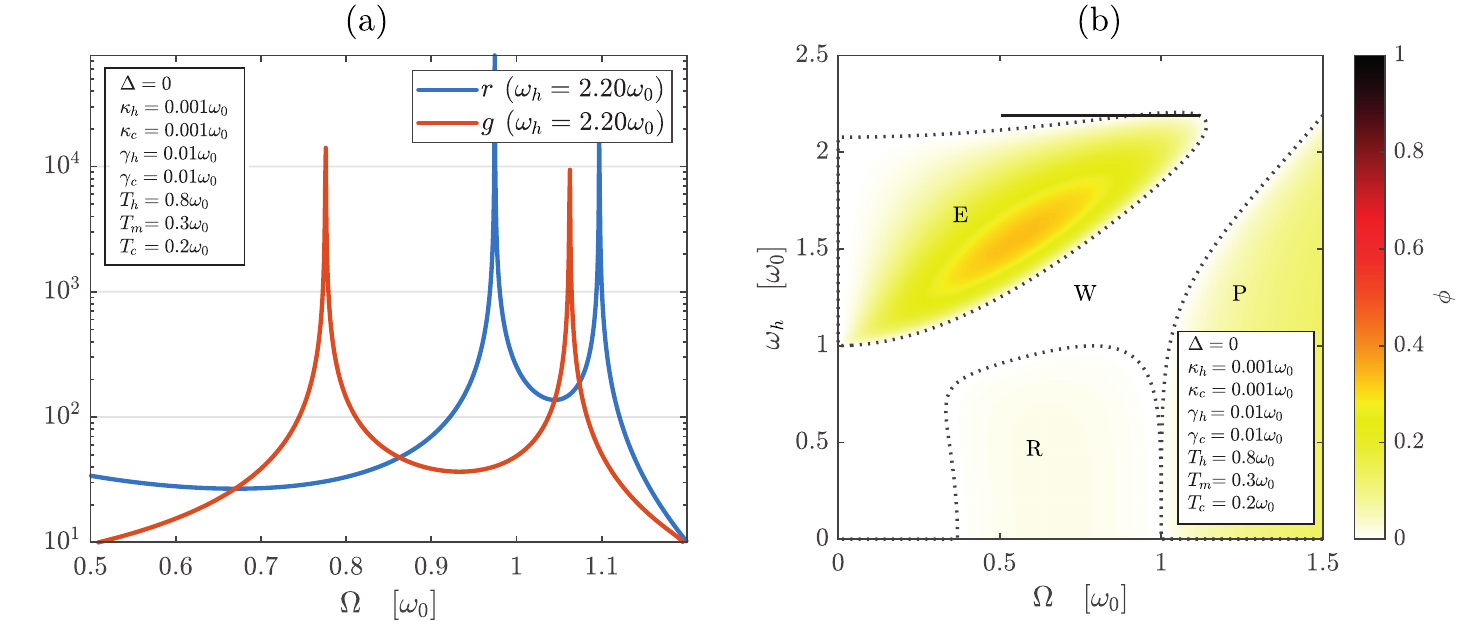}
  \caption{{\bf Three--terminal device as a quantum thermal transistor}. Panel (a) shows the output--to--input ratio $r$ and differential gain $g$ as a function of $\Omega$, for a set of parameters chosen in order to optimize the range of driving frequencies where the system performs as a thermal transistor. Panel (b) shows such range (as a thick solid line), superimposed to a density plot of the exergy efficiency $\phi$ as a function of $\Omega$ and $\omega_h$.}
  \label{fig:tranny} 
\end{figure}

We consider the exchanged power $P$, modulated through the driving frequency
$\Omega$ of the dynamical couplings, 
as the input, and the current $J_h$ flowing from the hot bath as the output. Similar results (not shown) can be obtained considering $J_c$ as the output. We introduce
\begin{equation}
    r=\left|\frac{J_h}{P}\right|\quad;\quad g=\left|\frac{\partial J_h}{\partial P}\right|=\left|\frac{\partial J_h/\partial\Omega}{\partial P/\partial\Omega}\right|\,.\label{eq:tranny}
\end{equation}
The parameter $r$ controls the output--to--input ratio, i.e. the "smallness" of the input signal with respect to the output, while $g$ represents the differential gain of the transistor, i.e. the amplification factor of small changes in $P$, reflected on $J_h$. We chose to customarily set a performance threshold, by considering a "useful" transistor a device in which both $r>10$ and $g>10$ is achieved, and performed a search scanning the parameters space.
Figure~\ref{fig:tranny} (a) shows a typical case in which the three--terminal device indeed works as a quantum thermal transistor: both criteria are met, and a gain up to $\sim10^4$ can indeed be achieved. The performance is stable, with the region in $\Omega$ over which $r,g>10$ extending over a frequency range $\gtrsim 0.5\omega_0$. We have found (not shown) that in order to achieve solid and robust transistor action, the best condition occurs with minimal detuning of the two Lorentzian baths $\Delta\approx 0$, with $T_m\approx T_c$. To put this result in context with the different operating modes of the device, Fig.~\ref{fig:tranny} (b) shows the region in parameter space where the device operates as a transistor as a solid black segment, superimposed to a plot of $\phi$ where the different operating modes are marked. The best transistor regimes are found near the boundaries between the "re--entrant" wasteful--engine--wasteful operating modes, where $P$ switches sign two times (see Fig.~\ref{fig:modes}) while $J_h$ preserves it. This clearly drives $r$ to very large values near the inversion points of $P$. It is found that in these regions $P$ is also sufficiently smooth with respect to $\Omega$, which in turn allows to obtain a large gain $g$. With no detuning, the exergy efficiency $\phi$ is found to be weak, reaching at most 0.4 (in contrast with Fig.~\ref{fig:multimode}, where $\Delta\neq 0$ and $\phi$ can approach the ideal unit efficiency). However, in the case of a thermal transistor a trade--off between $\phi$ and $r$,$g$ seems acceptable since in this regime the device is not intended to perform as a conventional thermal machine.

In our setup, the performance of a three--terminal transistor is far superior to that of a two--terminal device operating in the same manner and within the same temperature range -- obtained for instance switching off the coupling $g_c(t)=0$ -- for which the figures of merit $r$ and $g$ can be defined as in Eq.~(\ref{eq:tranny}). The results for this two--terminal transistor, shown in Fig.~\ref{fig:app1} (d) of Appendix~\ref{app:2}, clearly show that the driving range where a significant gain is achieved is much narrower, extending up to a frequency window of about $0.15\omega_0$: operating with three terminals warrants a definite advantage in terms of the stability of the thermal transistor.

In order to confirm the validity of choosing $\Omega$ as a modulation parameter, we have also analyzed the situation in which the transistor operates with $P$ as the input and $J_h$ as the output, but keeping $\Omega$ fixed and modulating the temperature $T_m$. In this case we have observed (not shown) that lower values for both $r$ and $g$ are obtained, in narrow regions of the temperature $T_m$.

\section{Discussion and Conclusions}
\label{sec:conclusions}

We have shown that by properly engineering bath spectral features 
and driving the system-bath couplings, a three-terminal quantum thermal machine
can act as an efficient hybrid thermal machine. The proposed device is flexible and 
tunable, in that several pure and hybrid working modes can be obtained, and 
it is possible to switch from one operational mode to another by changing 
the driving frequency. Moreover, the same device can work as a stable, high-performance transistor, with large output--to--input signal and 
differential gain, on a broad driving frequency interval. 

Natural extensions of our model could be obtained by considering hybrid 
quantum thermal machines with more complex working medium, like 
coupled oscillators or qubit-oscillator systems.
It would be also interesting to consider the coupling to 
nonequilibrium reservoirs,
which may be a useful resource for further boosting the performance of 
hybrid quantum thermal machines. Finally our analysis reported average 
quantities, while fluctuations, ubiquitous at the nanoscale, 
can influence device performance, but also provide alternative pathway 
in developing new device functionalities. In particular, 
the so-called thermodynamic uncertainty
relations, constraining in classical stochastic thermodynamics
the achievable degree of precision (i.e., the minimum 
fluctuations in the outcome of a process),
given a certain output power and efficiency~\cite{seifert18,horowitz19},
deserve further careful inspection in the quantum realm.

\section*{Acknowledgements}
F.C. and M.S. acknowledge support by the “Dipartimento di Eccellenza MIUR 2018-2022.”, L.R. and G.B. acknowledge financial support by the Julian Schwinger Foundation (Grant JSF-21-04-0001) and by INFN through the project ‘QUANTUM’. F.C. acknowledges support by V. Cavaliere for the preparation of Fig. 1.

\section*{Author Contributions}
F.C., L.R., and M.C. performed analytical calculations and numerical simulations.
M.S. conceived and supervised the study, with inputs from G.B.. 
All authors discussed the results and contributed to writing and revising the manuscript.

\section*{Declaration of interest}
The authors declare no competing interests.

\section*{Supplemental information}

\appendix

\section{Thermal currents and power in the perturbative regime}
\label{app:1}
The driven dissipative three-terminal setup described in the main text can be solved by resorting to non-equilibrium Green function formalism, as described in Ref.~\cite{carrega_prxquantum}. Here, we briefly recall the main steps for the evaluation of thermodynamic quantities of interest, referring to previous literature for additional details. From the total Hamiltonian in Eq.~(\ref{htot}) one can derive the equations of motion for the WM variables ($x(t)$, $p(t)$) and baths variables ($X_{k,\nu}(t)$, $P_{k,\nu}(t)$). The solution for the latter degrees of fredom can be expressed in terms of initial conditions and of the operator $x(t)$~\cite{carrega_prxquantum, zherbe95, paz1, paz2}, eventually obtaining a generalized quantum Langevin equation of the form
\begin{equation}\label{eq:diffqoper}
\ddot{x}(t) + \omega^2_0 x(t)+\int_{-\infty}^{+\infty}\!\!\mathrm{d}s\sum_{\nu=h,m,c} g_\nu(t)\gamma_\nu(t-s)\Big[\dot{g}_\nu(s)x(s)+
\dot{x}(s)g_\nu(s)\Big] =\frac{1}{M}\sum_{\nu=h,m,c} g_\nu(t)\xi_\nu(t)\,,
\end{equation}
where the overdots denote time derivatives,  we recall that $g_m(t)\equiv1$, the damping kernels $\gamma_{\nu}(t)$ are linked to the spectral function by
\begin{equation}\label{eq:gamma}
\gamma_\nu(t)=\frac{2}{\pi M}\vartheta(t)\int_0^{\infty}\mathrm{d}\omega \frac{{\cal J}_\nu(\omega)}{\omega}\cos(\omega t),
\end{equation} 
and $\xi_\nu (t)$ represent the fluctuating force operators of the baths with $\vartheta(t)$ the Heaviside step function. We recall that these operators have zero quantum average $\langle \xi_\nu(t)\rangle\equiv {\rm Tr}[\xi_\nu (t) \rho(t_0)]=0$ and their time correlators are given by $\langle \xi_\nu (t)\xi_{\nu'}(t')\rangle =\delta_{\nu,\nu'}{\cal L}_\nu (t-t')$, with
\be
\label{correlator}
\!\!\!\!\!\!{\cal L}_\nu (t)\!\!=\!\!\!\int_0^{\infty}\!\!\frac{\mathrm{d}\omega}{\pi} 
{\cal J}_\nu(\omega)\big[\!\coth(\frac{\omega}{2T_\nu})\!\cos(\omega t)-\!i \sin(\omega t)\big].
\ee
At long times, the system reaches a periodic state sustained by the drives. In this regime, the time evolution of the position operator $x(t)$ can be expressed directly as a time integral of the retarded Green function with  the inhomogeneous term:
\be
\label{solution}
x(t)=\sum_{\nu=h,m,c}\int_{-\infty}^{+\infty}\mathrm{d}t' G(t,t')\frac{1}{M}g_\nu(t')\xi_\nu(t').
\ee
Several methods can be used to evaluate (exactly or with different approximation schemes) the above Green function, see e.g. Refs.~\cite{carrega_prxquantum, cavaliere_prr, paz1, arrachea12}.\\
From Eqs.~(\ref{app_PSP}-\ref{app_JSP}), it is possible to express both the power and the heat currents averaged over one period ${\cal T}$ in terms of position variables as
\begin{equation}
\label{Pexpression}
P=\int_{\bar{t}}^{\bar{t}+{\cal T}}\frac{dt'}{{\cal T}}\Big\{\sum_{\nu=h,c} -\dot g_\nu(t')\langle x(t')\xi_\nu(t')\rangle+M\dot g_\nu(t')\int_{-\infty}^{+\infty}\!\!\!\!\!\!\!\mathrm{d}s\gamma_\nu(t'-s)\frac{{\mathrm d}}{{\mathrm d}s}
\left[g_\nu(s)\langle x(t')x(s)\rangle\right]\Big\}
\end{equation}
and
\begin{equation}
\label{heatexpression}
J_{\nu}=\int_{\bar{t}}^{\bar{t}+{\cal T}}\frac{d t'}{{\cal T}}\Big[-\frac{g_\nu(t')}{2}\left\langle x(t')\dot\xi_\nu(t')+\dot\xi_\nu(t')x(t')\right\rangle-g_\nu(t')\int_{-\infty}^{t'}\!\!\!\!\!\!\!\mathrm{d}s 
\left\langle \{x(t'),x(s)\}\right\rangle g_\nu(s)\int_0^{\infty}\!\!\frac{\mathrm{d}\omega}{\pi} 
{\cal J}_\nu(\omega)\omega\cos(\omega(t'-s)) \Big]\,,\nonumber
\end{equation}
with $\{A,B\}=AB+BA$ the standard anti-commutator.
It is worth to note that the average power gets contributions {\it only} from the two driven couplings $\nu=h,c$, since $\nu=m$ is kept constant with $g_m(t)=1$.\\
In this work we are interested in the weak coupling regime, where the driven bath couplings ($\nu=h,c$) are much weaker than the static one ($\nu=m$). In this case, closed analytical expressions for average power and heat currents can be obtained in a standard perturbative framework~\cite{cavaliere_prr}. In the regime $\gamma_m\ll\omega_0$, the final expressions read (for $\nu=h,c$)
\begin{equation}
J_{\nu}=\frac{1}{4M\omega_0}\sum_{p=\pm 1}(\omega_0+p\Omega)\mathcal{J}_{\nu}(\omega_0+p\Omega)\left[n_B\left(\frac{\omega_0+p\Omega}{T_{\nu}}\right)-n_B\left(\frac{\omega_0}{T_m}\right)\right]\,,\label{eq:curpert}
\end{equation}
\begin{equation}
P=-\frac{\Omega}{4M\omega_0}\sum_{\nu=h,c}\sum_{p=\pm
1}p\mathcal{J}_{\nu}(\omega_0+p\Omega)\left[n_B\left(\frac{\omega_0+p\Omega}{T_{\nu}}\right)-n_B\left(\frac{\omega_0}{T_m}\right)\right]\,,\label{eq:curpow}
\end{equation}
where $n_B(x)=\left(e^x-1\right)^{-1}$ is the Bose function and we note that Eqs.~(\ref{eq:curpert}),(\ref{eq:curpow}) do not depend anymore on $\gamma_m$ for $\gamma_m\ll\omega_0$~\cite{cavaliere_prr}. We do not quote the expression for $J_m$ as it can be derived from the conservation of energy in Eq.~(\ref{firstlaw}).

\section{Operating modes and thermal transistor effect in a two--terminal device}
\label{app:2}

\begin{figure}[h]
\includegraphics[width=0.8\linewidth]{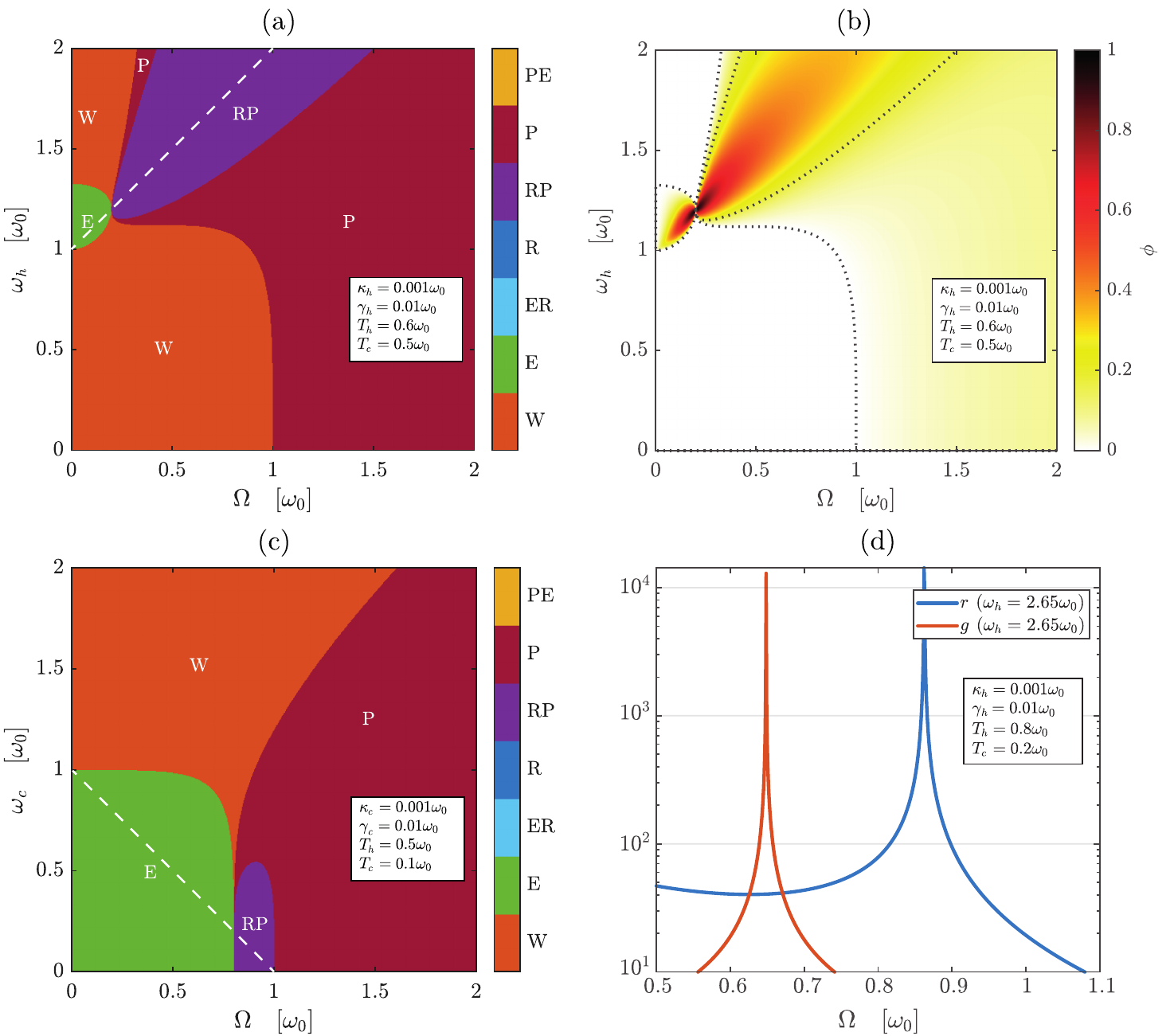}
  \caption{{\bf Two--terminal thermal machine}. Panel (a) shows the only four operating modes of a two--terminal thermal machine for the case of a Lorentzian hot bath. Panel (b) is a density plot of the exergy efficiency $\phi$ as a function of $\Omega$ and $\omega_h$ for the same parameters as in (a). Panel (c) displays the four operating modes for a two--terminal maghine with a Lorentzian cold bath. Panel (d) shows the output--to--input ratio $r$ and the gain $g$ as a function of $\Omega$, for a two--terminal machine operating as a thermal transistor with parameters similar to those in Fig.~\ref{fig:tranny}. In Panel (a) the white dashed lines represent the resonance $\omega_h=\omega_0+\Omega$, in Panel (c) the resonance $\omega_c=\omega_0-\Omega$.}
  \label{fig:app1} 
\end{figure}

In this section we briefly discuss the operation of a two--terminal quantum thermal machine with a QHO as the WM, operating between a Lorentzian bath with dynamical coupling and a static Ohmic bath~\cite{cavaliere_prr}. The theoretical framework can be derived from the material in Sec.~\ref{sec:model}, for vanishing coupling to one of 
the two Lorentzian baths.\\
The operating modes of a two--terminal device can be characterized with the same criteria shown in Fig.~\ref{fig:modes}. We denote the hot and cold baths as $T_h$ and $T_c<T_h$. A general landscape of {\em all} the possible operating modes is shown in Fig.~\ref{fig:app1} where panel (a) displays the case where the Lorentzian bath is at temperature $T_h$, and panel (c) illustrates the case of a Lorentzian bath at temperature $T_c$. These two cases correspond respectively to the machines $\mathcal{M}_1$ and $\mathcal{M}_2$ discussed in Sec.~\ref{sec:mmode}.\\
The only achievable modes here are the two pure "engine" and "heat pump" modes, the hybrid "refrigerator--pump" mode, and the "wasteful" mode. In passing, we note that other works~\cite{cavaliere_prr,Buffoni19,Solfanelli20} use different names for such operating modes: there, the terms "refrigerator", "dissipator" and "accelerator" correspond to "refrigerator--pump", "heat pump" and "wasteful" used here. Due to the presence of a sharply peaked Lorentzian bath, the machine displays its most prominent features when the resonance conditions $\omega_h=\omega_0+\Omega$ (for the machine $\mathcal{M}_1$) or $\omega_c=\omega_0-\Omega$ (for the machine $\mathcal{M}_2$) are met~\cite{cavaliere_prr}. These resonances are marked as a white dashed line in panels~(a),(c). The presence of such resonances is quite evident in the behaviour of the exergy efficiency, shown in Fig.~\ref{fig:app1}(b) for the case of $\mathcal{M}_1$ (the situation for $\mathcal{M}_2$ is qualitatively identical). Note that in the engine and refrigerator--pump regimes $\phi$ reduces respectively to the engine efficiency $\eta=-P/J_h$ normalized to the Carnot limit $\eta_C=1-T_c/T_h$, or to the coefficient of performance (COP) $\epsilon=J_c/P$ normalized to the Carnot limit $\epsilon_C=T_c/(T_h-T_c)$. The highest values of $\phi$ occur near the resonance line, with $\phi\to 1$ near the intersection between the resonance line and the line separating the engine and the refrigerator--pump regimes, where $P\to 0$.\\
Finally, Fig.~\ref{fig:app1}(d) shows the typical best--case scenario concerning the performances of a thermal transistor built out of a two--terminal device operating within a temperature range analogous to that of Fig.~\ref{fig:tranny}: clearly the range of driving frequencies where a useful gain $g>10$ is achieved is only of about $0.15\omega_0$, considerably smaller than the one obtained in the three--terminal case. This general trend has always been observed throughout all our investigations.

\begin{thebibliography}{90}
\bibitem{noteOhmic} Note that for strictly Ohmic regime it is assumed the usual bath cut-off $\omega_c$ as the largest energy scale, i.e. $\omega_c \gg \omega_0, \Omega$.
\end{thebibliography}
\end{document}